\documentclass[reprint, amsmath,amssymb, showkeys, aps]{revtex4-1}

\usepackage{graphicx}
\usepackage{dcolumn}
\usepackage{bm}

\newcommand{\Comment}[1]{{}}
\usepackage{amsfonts,amsthm,amsmath,amssymb,slashed}
\usepackage[textwidth = 500 pt, textheight = 630 pt]{geometry}

\Comment{\usepackage{color}
\definecolor{MyDarkBlue}{rgb}{0.15,0.15,0.45}
\usepackage[linktocpage=true]{hyperref}
\hypersetup{
colorlinks=true,
citecolor=MyDarkBlue,\bibliography{myref}
linkcolor=MyDarkBlue,
urlcolor=MyDarkBlue,
pdfauthor={Prieslei Goulart},
pdftitle={DRAFT},
pdfsubject={hep-th}
}

\usepackage[numbers,sort&compress]{natbib}
\usepackage{hypernat}}
\usepackage{graphicx}

\newcommand\ignore[1]{}
\def\one{{\,\hbox{1\kern-.8mm l}}}

\newcommand{\Cset}{{\,\,{{{^{_{\pmb{\mid}}}}\kern-.45em{\mathrm C}}}}}

\newcommand{\be}{\begin{equation}}
\newcommand{\bea}{\begin{eqnarray}}

\newcommand{\ee}{\end{equation}}
\newcommand{\eea}{\end{eqnarray}}

\usepackage{color}
\newcommand{\bse}{\begin{subequations}}
\newcommand{\ese}{\end{subequations}}

\begin{document}

\title{Massless black holes and charged wormholes in string theory}

\author{Prieslei Goulart}
\email{prieslei@ift.unesp.br}
 
\affiliation{
Instituto de F\'{i}sica Te\'{o}rica, UNESP-Universidade Estadual Paulista\\
R. Dr. Bento T. Ferraz 271, Bl. II, S\~{a}o Paulo 01140-070, SP, Brazil
}

\date{\today}

\begin{abstract}
We discuss the zero mass pointlike solutions and charged Einstein-Rosen bridges (wormholes) that arise from the dyonic black hole solution of the Einstein-Maxwell-dilaton theory. These massless black holes exist individually in spacetime, different from the known massless solutions, which come in pairs with opposite signs for their masses. In order to construct a massless object, we choose the integration constants of the solution to have specific values. The massless solutions present some problems: in one case the dilaton field is complex (or the gauge field has negative kinetic energy), and in the other case the solution has negative entropy and temperature or it is naked singularity in the extremal limit. For the first case, the observables computed are real quantities. This massless solution also allow the bridge construction, and we obtain an  analytical and static charged wormhole solution, which satisfies the null energy condition.

\end{abstract}

\keywords{Massless black holes, wormholes.}

\maketitle

\section{Introduction}
Zero mass black holes called a lot of attention after Strominger proposed that they were needed to explain the conifold singularities in the low energy theory describing the moduli space of Calabi-Yau vacua of type II string theory \cite{Strominger:1995cz}. Black holes in string theory with zero ADM mass exist \cite{Behrndt:1995tr,Kallosh:1995yz,Cvetic:1995mx,Ortin:1996nd}. These ones are solutions to the low energy effective theory, but their relation to the massless black holes proposed by Strominger is still unclear, since the latter arise at the conifold singularities, exactly when the low energy theory is singular and the semiclassical description breaks down. 

The solution in \cite{Behrndt:1995tr} is stable, since they have the minimal mass allowed by supersymmetry. All the others \cite{Kallosh:1995yz,Cvetic:1995mx,Ortin:1996nd,Emparan:1996am} are actually obtained by considering composite supersymmetric pairs (or several pairs) of extremal black holes with masses with "opposite signs" \cite{Ortin:1996fj}. The classical and quantum instabilities of these pairs are healed by coupling the theory with several scalar fields, which interact with their own scalar charges. These composite objects with masses having opposite signs provide a model for massless black holes, but their relation to the solution of \cite{Behrndt:1995tr} is still unknown. Moreover, there is no way to achieve zero mass limit of known non-extremal individual black holes. This violates the cosmic censor or introduces singular fields, as is the case of the Reissner-Nordstr\"{o}m (RN) or the dyonic solution for the Einstein-Maxwell-dilaton (EMD) solution \cite{Kallosh:1992ii}. 

The massless black holes are not the only misterious objects in the theory of general relativity. In an attempt to construct a geometrical model for an elementary particle which excludes sigularities, Einstein and Rosen \cite{Einstein:1935tc} (see also \cite{Visser:1995cc}), introduced the concept of a bridge: a region in spacetime that connects two separate sheets. A simple coordinate transformation that is valid everywhere outside the event horizon of a Schwarzschild black hole is enough to bring the metric into the form of a bridge. In modern terminology, the bridge is referred to as a {\it{wormhole}}. But the Schwarzschild wormhole, sometimes called neutral wormhole due to the absence of electromagnetic charges, is dynamical \cite{Fuller:1962zza}, and it pinches off so fast that not even light can cross the throat separating the two sheets. In other words, the neutral bridge is a non-traversable wormhole. Einstein and Rosen \cite{Einstein:1935tc} also tried to construct a charged bridge by using the RN black hole, as will be shown later, but this violated the null energy condition (NEC) everywhere in the spacetime. Matter that violates the NEC is called exotic matter, and there is no known charged static wormhole solution that does not require exotic matter to exist.  

In 1988, Morris and Thorne \citep{Morris:1988cz} investigated traversable wormholes, which are bridges that would allow a human being to cross them. Although quantum mechanically the energy is allowed to acquire negative values at the throat of a wormhole, the traversable solutions \cite{Friedman:1993ty,Hochberg:1998ii} together with the charged bridge discussed above require the existence of exotic matter for them to exist classically. The need of exotic matter can be circumvented by considering modified theories of gravity, see \cite{Kanti:2011jz,Bronnikov:2002rn,Sushkov:2011jh,Sushkov:2015fma,Ayon-Beato:2015eca} for instance. The solution of reference \cite{Ayon-Beato:2015eca} is a stationary traversable AdS wormholes with NUT parameter, and it is stated by the authors that a negative cosmological constant can "hardly" be considered exotic.  A possible type of exotic matter can be a phantom field, which is scalar field with negative kinetic energy \cite{Bronnikov:1973fh,Ellis:1973yv}. Gibbons and Rasheed \cite{Gibbons:1996pd} investigated the relation between these phantom fields and zero mass objects, such as black holes and wormholes. They discussed Dyson pairs production in the context of gravity coupled to vector fields and scalars, and showed that they consist of zero rest mass black holes with regular horizons. Another interesting and more recent idea, known as ER=EPR \cite{Maldacena:2013xja}, suggests that entangled black holes are connected by wormholes. 

In this paper we study the zero mass black holes  of EMD theory. We discuss how to obtain such massless objects by choosing specific values for the integration constants of the dyonic black hole solution of the EMD theory presented in \cite{Goulart:2016cuv}. Unfortunately, the non-extremal dyonic solution require singular dilaton field at infinity, although there are intriguing facts about it: The observables are physical quantities, and we such the massless black hole to construct static a charged wormhole, and show that this satisfies the NEC. We will discuss this in detail.

\section{Dyonic black hole for EMD theory}

We consider the EMD theory without a dilaton potential. The coupling to the field strength is inspired in supergravity models and take the usual exponential form (for the motivation to consider this theory, see \cite{Kallosh:1992ii} and references therein). The action is written as
\be S=\int d^{4}x\sqrt{-g}\left(R-2\partial_{\mu}\phi\partial^{\mu}\phi-e^{-2\phi}F_{\mu\nu}F^{\mu\nu}\right). \label{ad}\ee
We take units in which $(16\pi G_N)\equiv 1$, where $G_N$ is the Newton's constant. The field strength has the usual definition, i.e.
\be F_{\mu\nu}=\partial_{\mu}A_{\nu}-\partial_{\nu}A_{\mu}. \ee
The dyonic black hole solution written in terms of integration constants takes the form \cite{Goulart:2016cuv}
\begin{align}
ds^{2}&=-e^{-\lambda}dt^{2}+e^{\lambda}dr^{2}+C^{2}(r)d\Omega^{2}_{2}, \label{genmets} \\
e^{-\lambda}&= \frac{(r-r_{1})(r-r_{2})}{(r+d_{0})(r+d_{1})}, \,\,\, C^{2}(r)=(r+d_{0})(r+d_{1}),\label{metelements}\\
e^{2\phi}&=e^{2\phi_{0}}\frac{r+d_{1}}{r+d_{0}}, \label{gendil} \\
F_{rt}&=\frac{e^{2\phi_{0}}Q}{(r+d_{0})^{2}}, \,\,\, F_{\theta\phi}=P\sin\theta. \label{gauge}\end{align}
The parameters o the solution are: the electic charge $Q$, the magnetic charge $P$, the value of the dilaton at infinity $\phi_{0}$, and four integration constants, $r_{1}$, $r_{2}$, $d_{0}$ and $d_{1}$. This solution is totally free of boundary conditions. The equations of motion imply that the parameters must respect the relations
\be (d_{0}-d_{1})[(r_{1}+r_{2})+(d_{0}+d_{1})]=2(e^{2\phi_{0}}Q^{2}-e^{-2\phi_{0}}P^{2}), \label{eq1}\ee
\be (d_{0}-d_{1})(d_{0}d_{1}-r_{1}r_{2})=2(d_{1}e^{2\phi_{0}}Q^{2}-d_{0}e^{-2\phi_{0}}P^{2}),  \label{eq2}\ee
\bea (d_{0}-d_{1})[-(r_{1}+r_{2})d_{0}d_{1}-(d_{0}+d_{1})r_{1}r_{2}]\nonumber \\
=2(d_{1}^{2}e^{2\phi_{0}}Q^{2}-d_{0}^{2}e^{-2\phi_{0}}P^{2}). \label{eq3}\eea
Some physical quantities are defined in terms of the integration constants. The dilaton charge, for instance, is defined as
\be \Sigma=\frac{1}{4\pi}\int d\Sigma^{\mu}\nabla_{\mu}\phi=\frac{(d_{0}-d_{1})}{2}. \label{chargedilaton}\ee
Notice that the $g_{tt}$ component of the metric have the following expansion far from the black hole
\be g_{tt}=-\left(1-\frac{(d_{0}+d_{1}+r_{1}+r_{2})}{r}\right)+\mathcal{O}\left(\frac{1}{r^{2}}\right). \ee
In the Newtonian approximation the mass $M$ of the black hole is given by 
\be (d_{0}+d_{1}+r_{1}+r_{2})=2M. \label{defmass}\ee
As was discussed in \cite{Goulart:2016cuv}, one can solve (\ref{eq1}), (\ref{eq2}), and (\ref{eq3}) and obtain $r_{1}$ and $r_{2}$ in terms of the other constants. Using the definitions of physical charges (\ref{chargedilaton}) and (\ref{defmass}) one can recover the dyonic solution found by Kallosh et al. \cite{Kallosh:1992ii}, which is a well-defined solution in the limit when the dilaton charge is zero, but not when the mass is zero. We now show that having the solution written in terms of the integration constants allows us to construct massless solutions. 

\section{Massless pointlike dyonic solutions}
Notice that (\ref{eq1}) has the definition of mass (\ref{defmass}) in it. In order to obtain $M=0$, we set $d_{1}=-d_{0}=- \Sigma$ and $r_{1}=-r_{2}\equiv r_{H}$ at the same time, and this implies that directly that
\be e^{2\phi_{0}}=\pm\frac{P}{Q}.\label{dilinfty} \ee 
We will discuss the situations corresponding to each sign. Equation (\ref{eq2}) fixes the event horizons as
\be r_{H}^{2}= \Sigma^{2}\mp 2QP. \label{hor}\ee
Notice that the positive sign in (\ref{dilinfty}) implies that we must take the minus sign in (\ref{hor}). This also respects (\ref{eq3}), which shows consistency with all the equations motion. 
\begin{itemize}
\item $e^{2\phi_{0}}=-\frac{P}{Q}$
\end{itemize}
We choose the minus sign in (\ref{dilinfty}), and discuss the physical relevance of this choice later. The non-extremal solution is written as 
\begin{align} 
ds^{2}&=-e^{-\lambda}dt^{2}+e^{\lambda}dr^{2}+C^{2}(r)d\Omega^{2}_{2},  \nonumber \\
e^{-\lambda}&=\frac{(r-r_{+})(r-r_{-})}{(r^2-\Sigma^2)}, \,\,\, C^{2}(r)=(r^2-\Sigma^2),  \label{nonextmet-} \\
e^{2\phi}&=-\frac{P}{Q}\frac{(r-\Sigma)}{(r+\Sigma)}, \label{dil-} \\
F_{rt}&=-\frac{P}{(r+\Sigma)^{2}}, \,\,\, F_{\theta\phi}=P\sin\theta. \label{gauge-}
\end{align} 
The horizon and singularity are located at 
\be r_{+}=+ \sqrt{\Sigma^{2}+ 2QP}, \,\,\, r_{S}=|\Sigma |.  \ee
Notice that the area of the two-sphere shrinks to zero at $r_{S}$. This excludes $r_{-}=- \sqrt{\Sigma^{2}+ 2QP}$ as an inner horizon, since the angular part of the metric will flip sign when an observer approaches this region, leading to problems with causality. The temperature $T$ and entropy $S$ associated to this object are given by
\be T=\frac{1}{4\pi}\frac{\sqrt{\Sigma^{2}+ 2QP}}{ 2QP}, \,\, S= 2\pi QP.\label{tszeromass}\ee
The temperature and the entropy are positive quantities. Notice also that the dyonic massless black hole can not become extremal, since this implies that we have an imaginary dilaton charge. The entropy depends only on the electric and magnetic charges, and has the same value as the entropy of extremal black holes of EMD theory with arbitrary dilaton charge and non-zero mass. When the dilaton charge is zero, we still have a non-extremal black hole, with horizon at $r_{+}=+ \sqrt{2QP}$. This shows that, at the critical point of the moduli space, i.e. $\Sigma=0$, we indeed have a massless black hole solution, which is non-extremal and have temperature and entropy given by (\ref{tszeromass}).

We see that it is indeed possible to construct a massless dyonic black hole solution for the EMD theory. In order to do that, we had to fix the dilaton field at infinity to be complex. The minus sign in (\ref{dilinfty}) seems to spoil this solution, although all the physical quantities are real. Zero mass electrically charged solutions of EMD theory were discussed by Gibbons and Rasheed in \cite{Gibbons:1996pd}. These authors obtained massless solutions for such a theory by flipping the sign of the kinetic term of the dilaton or the gauge field, or of both terms at the same time, introducing the term "anti" to express which kinetic term has a flipped sign. The Einstein-Maxwell-anti-Dilaton (EMaD) theory for instance, has a positive kinetic term in the action for the dilaton, and so on. The motivation for doing so was based on the Dyson's argument \cite{Gibbons:1996pd}: The properties of the theory 
\be -\frac{1}{4e^{2}}F_{\mu\nu}F^{\mu\nu} \ee
do not depend analytically on the coupling constant $e^{2}$. If this was not the case, then perturbation theory around the origin would be convergent in powers of $e^{2}$, and also in powers of $-e^{2}$. For the negative sign, the particles would attract, destabilizing the vacuum. As extremal charged black holes behave as charged particles, the authors of \cite{Gibbons:1996pd} used Dyson's argument \cite{Dyson:1952tj} to study what happens to black holes in theories with positive kinetic terms. In fact, they construct massless electrically charged black holes and wormholes for such theories. Here, studying the dyonic case, we did not flip the sign of any kinetic term, but the massless solution introduced, as stated above, a complex dilaton field at infinity. One can check that it is possible to obtain the same massless solution with a real dilaton field if we flip the sign of the kinetic term for the gauge field, which was the same situation studied in \cite{Gibbons:1996pd} for the electrically charged case. So, by trying to avoid problems with the dilaton field we end up transfering the problem to the gauge field, making it have negative kinetic energy. This does not come as a surprise. As we will use this massless solution to build wormholes in the next section, it is worth pointing out that there are other examples in the literature in which wormholes exist when the fields have negative kinetic energy, or imaginary electromagnetic charges. Euclidean wormholes were shown to exist as solutions to low-energy effective actions in string theory, whether when the charges are imaginary \cite{Bergshoeff:2004pg}, or when one of the fields have negative kinetic energy \cite{ArkaniHamed:2007js}. Notice also that non-Abelian gauge fields with negative kinetic energies have zero mass monopoles, as was pointed out in reference \cite{Lindstrom:1995ej}. Our analysis shows that, even with a complex dilaton field at infinity, this masless solution seems physically acceptable, since the observables are real. Our intention is not to prove whether or not this is the case, but instead, to show that this solution can be used to construct Eintein-Rosen bridges. As the scalar field does not seem to be physical, one would expect that the NEC is not satisfied. We will also show that this is not the case: NEC is satisfied.

\begin{itemize}
\item $e^{2\phi_{0}}=+\frac{P}{Q}$
\end{itemize}
This other massless solution is written as
\begin{align} 
ds^{2}&=-e^{-\lambda}dt^{2}+e^{\lambda}dr^{2}+C^{2}(r)d\Omega^{2}_{2},  \nonumber \\
e^{-\lambda}&=\frac{(r-r_{+})(r-r_{-})}{(r^2-\Sigma^2)}, \,\,\, C^{2}(r)=(r^2-\Sigma^2),  \label{nonextmet+} \\
e^{2\phi}&=\frac{P}{Q}\frac{(r-\Sigma)}{(r+\Sigma)},\label{dil+} \\
F_{rt}&=\frac{P}{(r+\Sigma)^{2}}, \,\,\, F_{\theta\phi}=P\sin\theta. \label{gauge+}
\end{align} 
The singularity is located at 
\be r_{S}=|\Sigma |.  \ee
The quantities $r_{\pm}=\pm \sqrt{\Sigma^{2}- 2QP}$ are always smaller than $r_{S}$, so this solution represents a naked singularity. The problems related to the previous case are absent here, since the value of the dilaton field at infinity is a real quantity. But still, this massless solution does not have a horizon, and, as we will see, it can not be used for the bridge construction.

The massless solution in \citep{Behrndt:1995tr} is written as 
\be ds^{2}=-\left(1-\frac{D^{2}}{r^{2}}\right)^{-1/2}dt^{2}+\left(1-\frac{D^{2}}{r^{2}}\right)^{1/2}d\vec{x}^{2}. \ee
A question that is still unanswered is whether this solution represents the extremal limit of a known black hole solution. We see that this is not the case for the massless solutions obtained for the EMD theory. 

\section{Bridge construction}
Consider the RN solution
\begin{align}
ds^{2}&=-e^{-\lambda}dt^{2}+e^{\lambda}dr^{2}+r^{2}d\Omega^{2}_{2}, \nonumber \\
e^{-\lambda}&= 1-\frac{2M}{r}+\frac{Q^{2}+P^{2}}{r^{2}},\nonumber \\
F_{rt}&=\frac{Q}{r^{2}}, \,\, F_{\theta\phi}=P\sin\theta.\label{RNsol}
\end{align}
By setting the mass $M$ to zero, the solution turns into a naked singularity, i.e. it has no horizon. The existence of a horizon is a necessary condition for the bridge construction, so, Eintein and Rosen \cite{Einstein:1935tc} considered imaginary charges, such that the sign in front of the term containing the charges squared would be minus. This corresponds to setting $(Q^{2}+P^{2})\equiv -\epsilon^{2}$, for a constant $\epsilon$. This solution now has a horizon, and then the coordinate change $u^{2}=r^{2}-\epsilon^{2}$ brings the metric into the bridge form
\be ds^{2}=-\frac{u^{2}}{u^{2}+\epsilon^{2}}dt^{2}+du^{2}+(u^{2}+\epsilon^{2})d\Omega_{2}^{2}. \label{ERbridge}\ee
The throat of this charged bridge is at $u=0$. But the matter in this solution is exotic: it has negative energy density everywhere in space. 

We will see that this is not the case for the bridges constructed from massless non-extremal solution of the previous section. For the moment, we will take a more general situation in which $r_{1}=-r_{2}\equiv r_{0}$. Notice that, by switching to the coordinates $u^{2}=r^{2}-r_{0}^{2}$, the metric (\ref{genmets}) is written as
\begin{align}
 ds^{2}&=-\frac{u^{2}}{(u^{2}+r_{0}^{2})}\frac{dt^{2}}{f(u)}+f(u)du^{2}+ (u^{2}+r_{0}^{2})f(u)d\Omega_{2}^{2}, \nonumber \\
 f(u)&=\left(1+\frac{d_{0}}{\sqrt{u^{2}+r_{0}^{2}}}\right)\left(1+\frac{d_{1}}{\sqrt{u^{2}+r_{0}^{2}}}\right). \label{nonextwh}
\end{align}
This is a genuine charged wormhole solution: It connects one Minkowski space at $u=-\infty$ to another at $u=+\infty$. The throat of the wormhole is located at $u=0$, and it has radius 
\be R_{\text{throat}}=\sqrt{(r_{0}+d_{0})(r_{0}+d_{1})}, \ee
where $d_{0}$ and $d_{1}$ will be determined by equations (\ref{eq1}), (\ref{eq2}), and (\ref{eq3}). The term inside the square root is positive when we take the minus sign in (\ref{dilinfty}), and the throat of the wormhole will always be greater than zero for $|Q|,|P|>0$. Notice that this solution is valid only outside the horizon, $r> +r_{0}$. So, for the full massless non-extremal solution (\ref{nonextmet-}), we must take the minus sign in (\ref{dilinfty}), and $d_{1}=-d_{0}\equiv -\Sigma$. The solution is then
\bea
 ds^{2}=-\frac{u^{2}}{(u^{2}+2QP)}dt^{2}+\frac{u^{2}+2QP}{u^{2}+\Sigma^{2}+2QP}du^{2} \nonumber \\
+(u^{2}+2QP)d\Omega_{2}^{2}. \label{zeromasswh}
\eea
At the critical point of the moduli space, $\Sigma=0$, this bridge is exactly the Einstein-Rosen bridge (\ref{ERbridge}), with $\epsilon^{2}=2QP$. This is due to the fact that  $Q\propto -P$, which fulfills $\epsilon^{2}\propto -Q^{2}$. The radius of the throat is
\be R_{\text{throat}}=\sqrt{2QP}. \ee
We see that the charged wormholes in the EMD theory may come from the non-extremal massless dyonic solutions. 

\section{Null Energy condition}
Unlike the charged wormholes constructed from the RN solution, we did not need to consider imaginary charges to achieve (\ref{nonextwh}). This gives some hope that these charged wormholes do not require exotic matter to exist. As stated before, exotic matter violates the NEC, and we now check whether this is the case in the present paper or not. We will use the original $r$ coordinate, but the same analysis can be done using $u$ coordinate, since the null energy condition does not depend on the choice of coordinate system. We follow the same prescription as in \cite{Morris:1988cz}, and prove our statements for the non-extremal massless solution (\ref{nonextmet-}). Again, taking $r_{1}=-r_{2}\equiv r_{0}$, the Ricci tensors for the metric (\ref{genmets}) are given by
\begin{align}
 R_{tt}&=\frac{\left(r^2-r_{0}^2\right)}{2 (d_{0}+r)^4 (d_{1}+r)^4} \left[d_{0}^2 \left(2 d_{1}^2+2
   d_{1} r+r^2-r_{0}^2\right)\right. \nonumber\\
   &   +2 d_{0} r
   \left(d_{1}^2-r_{0}^2\right)+d_{1}^2
   \left(r^2-r_{0}^2\right)\left. -2 d_{1} r r_{0}^2-2 r^2
   r_{0}^2\right], \\
 R_{rr}&=\frac{r_{0}^2-d_{0} d_{1}}{(d_{0}+r) (d_{1}+r)
   \left(r^2-r_{0}^2\right)}, \\
 R_{\theta \theta}&=\frac{1}{2
   (d_{0}+r)^2 (d_{1}+r)^2}\left[d_{0}^2 \left(2 d_{1}^2+2 d_{1} r+r^2-r_{0}^2\right)\right.\nonumber \\
   & +2
   d_{0} r \left(d_{1}^2-r_{0}^2\right)+d_{1}^2
   \left(r^2-r_{0}^2\right)\left. -2 d_{1} r r_{0}^2-2 r^2 r_{0}^2\right], \\
R_{\phi\phi}&= R_{\theta \theta}\sin^{2}\theta. \end{align}
The curvature tensor is 
\be R=\frac{(d_{0}-d_{1})^2 \left(r^2-r_{0}^2\right)}{2 (d_{0}+r)^3
   (d_{1}+r)^3}.  \ee
We choose orthonormal basis vectors \cite{Morris:1988cz}:
\begin{align}
{\bf{e}}_{\hat{t}}&=\left(\frac{(r+d_{0})(r+d_{1})}{(r^{2}-r_{0}^{2})}\right)^{1/2}{\bf{e}}_{t},\\
{\bf{e}}_{\hat{r}}&=\left(\frac{(r^{2}-r_{0}^{2})}{(r+d_{0})(r+d_{1})}\right)^{1/2}{\bf{e}}_{r},\\
{\bf{e}}_{\hat{\theta}}&=\left(\frac{1}{(r+d_{0})(r+d_{1})}\right)^{1/2}{\bf{e}}_{\theta},\\
{\bf{e}}_{\hat{\phi}}&=\left(\frac{1}{(r+d_{0})(r+d_{1})}\right)^{1/2}\frac{1}{\sin\theta}{\bf{e}}_{\phi}.
\end{align}
In this basis the metric coefficients take the form ${\bf{g}}_{\hat{\alpha}\hat{\beta}}={\bf{e}}_{\hat{\alpha}}\cdot {\bf{e}}_{\hat{\beta}}={\bf{\eta}}_{\hat{\alpha}\hat{\beta}}=\text{diag}(-1,1,1,1)$. 
Einstein's equations take the form
\be G_{\hat{\mu}\hat{\nu}}=8\pi G_{N}T_{\hat{\mu}\hat{\nu}}.\ee
In our unities $(16 \pi G_{N})=1$. The components of the energy momentum tensor are  $T_{\hat{t}\hat{t}}=\rho(r)$, $T_{\hat{r}\hat{r}}=-\tau(r)$, $T_{\hat{\theta}\hat{\theta}}=T_{\hat{\phi}\hat{\phi}}
=p(r)$, where $\rho(r)$ is the energy density measured by the static observer, $\tau(r)$ is the tension per unit area measured in the radial direction, and $p(r)$ is the pressure that is measured in the directions orthogonal to the radial direction. They are given by
\begin{align} 
\rho(r) &=\frac{1}{2
   (d_{0}+r)^3 (d_{1}+r)^3}\left[ 4 d_{0} r
   \left(d_{1}^2-r_{0}^2\right) \right.\nonumber \\
   & 2 d_{0}^2 \left(2 d_{1}^2+2 d_{1}
   r+r^2-r_{0}^2\right)+2d_{1}^2
   \left(r^2-r_{0}^2\right)\nonumber\\
   &  \left. -4 d_{1} r r_{0}^2-4 r^2
   r_{0}^2+(d_{0}-d_{1})^2 \left(r^2-r_{0}^2\right)\right], \\
 -\tau(r) &=\frac{1}{2
   (d_{0}+r)^3 (d_{1}+r)^3}\left[-2(d_{0}-d_{1})^{2}(r^{2}-r_{0}^{2})\right. \nonumber\\
   &  \left. +(4r_{0}^{2}-4d_{0}d_{1})(d_{0}+r) (d_{1}+r)\right].
\end{align}
The NEC states that
\be T_{\hat{\mu}\hat{\nu}}k^{\hat{\mu}}k^{\hat{\nu}}\geq 0. \ee
In the same coordinate system, the null vector is given by $k^{\hat{\mu}}=(1,1,0,0)$, and the NEC results in
\be \rho(r)-\tau(r)=\frac{(d_{0}-d_{1})^{2}(r^{2}-r_{0}^{2})}{2
   (d_{0}+r)^3 (d_{1}+r)^3}.  \ee
Evaluating this for the metric (\ref{nonextmet-}), in which $e^{2\phi_{0}}=-P/Q$, and $d_{1}=-d_{0}\equiv -\Sigma$, we have
\be \rho(r_{0})-\tau(r_{0})=\frac{2\Sigma^2(r^{2}-(\Sigma^{2}+2QP))}{(r^{2}-\Sigma^{2})^{3}}\ge 0. \ee
This shows that the NEC is satisfied, and the charged wormhole solution presented here does not require exotic matter to exist. The coordinate system is valid only outside the horizon, and the only way to saturate the bound is at the critical point of the moduli space, i.e.  
\be \Sigma=0 \Rightarrow \rho(r_{0})-\tau(r_{0})= 0. \ee
The massless pointlike objects presented here are entirely new, and, of course, the charged wormholes may be understood as a generalization of the charged Einstein-Rosen bridge, for the case when we include the dilaton in the theory.

There are many directions in which one may follow to answer the questions that may have arisen in the text. The ADM mass for the solutions discussed here is zero. Negative mass is in general a sign of instability, but a full analysis is necessary to prove that our massless solutions are stable under small perturbations of the metric. A topic for a future work is whether or not the same argument concerning the traversability of the neutral bridges applies here. It is intriguing though that the massless black holes solution, with a complex dilaton field at infinity, allowed us to construct a wormhole that satisfies the NEC. As stated in the text, a complex dilaton field at infinity can be avoided by allowing the gauge fields to have negative kinetic energy. Quantum mechanically, energy is allowed to admit negative values, but we are dealing with a classical theory. In general, a classical theory admiting fields whose kinetic energy is negative violates the NEC, but here we just saw a counterexample of such claim. The physical observables computed here are the temperature, entropy, mass, electric charge, and magnetic charge, and all of them are real quantities. The dilaton charge does not depend on $\phi_{0}$, and this is also a real quantity. The electric charge appears in $F_{rt}$ as $e^{2\phi_{0}}Q$, and this is real, although $\phi_{0}$ is complex. These facts are strong indications that the solution (\ref{nonextmet-}), (\ref{dil-}), and (\ref{gauge-}) is indeed physically acceptable, but a more careful analysis is necessary in order to make such a claim. This is a topic of future research.

\section{Conclusions}
In this paper we presented the massless pointlike objects arising as solutions to the EMD theory. They can be a massless non-extremal black hole, or a naked singularity. This shows that massless black holes exist individually, without the need of coming in pairs. All the physical quantities comuputed for the non-extremal solution are real. From the non-extremal solution, we constructed a static charged bridge and showed that this satisfies the NEC.

\begin{acknowledgments}
The author is grateful to George Matsas, Horatiu Nastase and Pedro Vieira for useful discussions. The author is grateful for the hospitality of the Max-Planck-Institut f\"{u}r  Physik (Werner Heisenberg Institut), where part of this work was developed. This work is supported by FAPESP grant 2013/00140-7 and 2015/17441-5.
\end{acknowledgments}



%

\end{document}